\documentstyle[sprocl,epsfig]{article}
\begin{document}

\title{SOLVABLE RECTANGLE TRIANGLE RANDOM TILINGS}

\author{J. DE GIER, B. NIENHUIS}

\address{Insitute for Theoretical Physics, University of Amsterdam,
Valckenierstraat 65, 1018 XE Amsterdam, The Netherlands.}

\maketitle\abstracts{
We show that a rectangle triangle random tiling with a tenfold symmetric
phase is solvable by Bethe Ansatz. After the twelvefold square
triangle and the eightfold rectangle triangle random tiling, this is
the third example of a rectangle triangle tiling which is solvable. A Bethe
Ansatz solution provides in principle an accurate
estimate of the entropy and phason elastic constants. In the
twelvefold and eightfold cases even {\em exact} analytic expressions have been
obtained from the Bethe Ansatz solution.   
}

\section{Introduction}
A random tiling ensemble with a tenfold symmetric phase is defined by
rectangles and isosceles triangles of sides of length 1 and $l=2\cos
(3\pi/10)=\sqrt{2+\tau}/\tau$, where $\tau=(\sqrt{5}+1)/2$ is the
golden mean. This random tiling was used by He {\em et
al.}\,\cite{He:1990} and by Nissen and Beeli \cite{Nissen:1993} to model a
decagonal phase in FeNb, by Oxborrow and Mihalkovi\v c \cite{Oxborrow:1994}
to model disorder in decagonal AlPdMn and by Roth
and Henley \cite{Roth:1996} to model the equilibrium structure
resulting from a molecular dynamics simulation.  The perfect
quasiperiodic tiling corresponds to a maximum possible density of a
decagonal disc packing.\,\cite{Cockayne:1995}  
The aim of this work is to calculate the entropy of the random tiling
and its phason elastic constants. To perform this calculation we use a
transfer matrix that generates the ensemble of
tilings.\,\cite{Henley:1988} The logarithm of the largest eigenvalue of
this matrix will give the free energy in the thermodynamic limit.

\section{Bethe Ansatz}
\begin{figure}[h]
\centerline{\epsffile{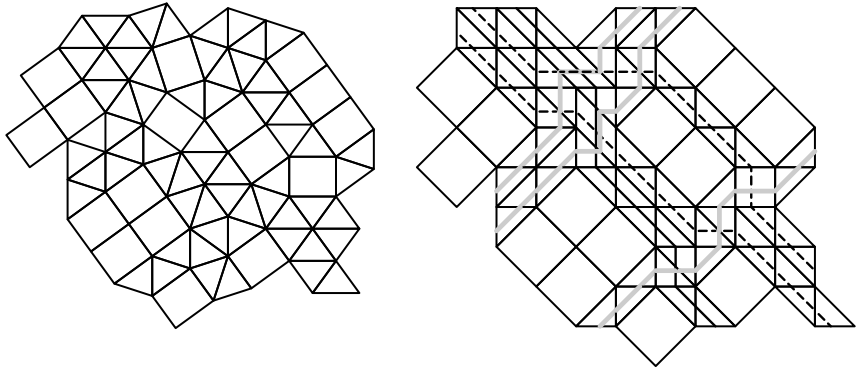}}
\caption{a) Patch of the random tiling. b) Corresponding patch on the
square lattice\label{fig:tiling}}
\end{figure}

We show that the model is solvable in the sense that its
transfer matrix can be diagonalized using Bethe Ansatz techniques.
First we deform the tiling such that its vertices are a subset of
those of the square lattice. Similar to the 
twelvefold square-triangle tiling \cite{Widom:1993,Kalugin:1994} and
the octagonal rectangle-triangle tiling,\,\cite{Gier:1996a} we have to
decorate the deformed tiles with lines to make this deformation map
bijective. An example of the deformation and decoration is given in
Fig.\ref{fig:tiling}. 
From this example it is seen that the triangles form domain walls
between patches of one type of rectangle. There are three types of
domain wall; one type is running from bottom-left to top-right (right
mover), denoted 
by the gray lines in Fig.\ref{fig:tiling}b, the other two are running
form bottom-right to top-left (left movers) and are denoted by the solid and
dashed lines in Fig.\ref{fig:tiling}b. These lines scatter with each
other, but they persist through the lattice. This means that the
number of each type of domain wall is a conserved quantity under the
action of the transfer matrix. We denote the numbers of gray, solid
and dashed lines by $m$, $n_1$ and $n_2$ respectively. A state
$\alpha$ on a row of the lattice can be specified by the positions
$y_1,\dots,y_m$ of the right movers and by the positions
$x_1,\ldots,x_n$ of all the left movers $n=n_1+n_2$ with the
specification that the 
lines $i_1,\ldots,i_{n_2}$ at positions $x_{i_2},\ldots,x_{i_{n_2}}$
are dashed lines. Elements $\psi(\alpha)$ of the eigenvector thus can be
written explicitly as $\psi(i_1,\dots,i_{n_2}|x_1,\ldots,x_n;y_1,\ldots,y_m)$. 
We make the following (plane wave) Ansatz for the form of the eigenvector
\begin{eqnarray}
\psi(i_1,\dots,i_{n_2}|x_1,\ldots,x_n;y_1,\ldots,y_m)=\hskip5cm\nonumber\\
\sum_{\pi,\rho} \sum_\mu A(\Gamma) B(\mu)
\prod_{a=1}^n z_{\pi_a}^{x_a} \prod_{b=1}^m w_{\rho_b}^{y_b}
\prod_{c=1}^{n_2} \prod_{r=1}^{i_c}
u(\mu_c,\pi_r). \label{eq:eigvec_ans} 
\end{eqnarray}
The amplitudes $A$ depend on the permutations $\pi$ and $\rho$ and on
the configuration of the various domain walls.\,\cite{Gier:1996a}
Simlarly, the amplitudes $B$ depend on the permutation $\mu$ and on
the sequence of dashed and solid black lines. 

If all the domain walls are separated the action of the transfer
matrix is just a shift of each line to the right or to the left.
The eigenvalue corresponding to the vector
(\ref{eq:eigvec_ans}) is therefore given by
\begin{equation}
\Lambda = \prod_{i=1}^n z_i \prod_{j=1}^m w_j^{-1}.\label{eq:eigval}
\end{equation}
At places where different domain walls are close together, the action
of the transfer matrix is not given by a mere shift of all domain
walls. These more complex processes turn out to be consistent with the
form (\ref{eq:eigvec_ans}) of the eigenvector and impose constraints on the
amplitudes $A$ and $B$ and the $z$'s, $w$'s and $u$'s of the eigenvector
(\ref{eq:eigvec_ans}). These equations are the socalled Bethe Ansatz
equations and are given by
\begin{eqnarray}
w_j^{-N} &=& (-)^{m-1} \prod_{k=1}^m \left( \frac{w_j}{w_k}\right)
\prod_{i=1}^n \left( z_i w_j + z_i^{-1} w_j^{-1}\right),\label{eq:BA1}\\
z_i^N &=& (-)^{n-1}\prod_{k=1}^n \left( \frac{z_k}{z_i}\right)
\prod_{j=1}^m \left( z_i w_j + z_i^{-1} w_j^{-1}\right)
\prod_{l=1}^{n_2} u(l,i),\label{eq:BA2}\\
(-)^{n_2-1} &=& \prod_{i=1}^n u(l,i),\label{eq:BA3}
\end{eqnarray}
where $u$ is given by
\begin{equation}
u(l,i) = \left(v_l+z_i^2-z_i^{-2}\right)^{-1}.\label{eq:BA4}
\end{equation}
The Ansatz (\ref{eq:eigvec_ans}) and the Eqs.
(\ref{eq:BA1})-(\ref{eq:BA4}) can be generalized to include chemical
potentials to control the tile densities in the ensemble.

\section{Entropy}
The entropy can now be calculated from the eigenvalue
(\ref{eq:eigval}). Some finite size results for the entropy per
lattice site and extrapolation are given in Table~\ref{tab:data}.
\begin{table}[h]
\caption{Numerical data for the entropy \label{tab:data}}
\vspace{0.4cm}
\begin{center}
\begin{tabular}{|c|l|}
\hline
$N$ & $\sigma_N$ \\ \hline
8  & 0.19482 \\
13 & 0.17779 \\
21 & 0.17088 \\
34 & 0.16853 \\
55 & 0.16763 \\
89 & 0.16730 \\ \hline
$\infty$ & 0.1671 (1) \\\hline
\end{tabular}
\end{center}
\end{table}

According to the random tiling hypothesis the entropy per area has its
maximum at the point where the densities match those of the
quasiperiodic tiling. For the tenfold tiling the expression for the
entropy in terms of the phason strain near the quasicrystalline point
is given by
\begin{eqnarray}
\sigma_a({\bf E}) &=& \sigma_0 - \frac{1}{4} K_1 \left((E_{1x} + E_{2y})^2 +
(E_{1y} - E_{2x})^2\right)\nonumber\\
&& {} - \frac{1}{4} K_2 \left((E_{1x} - E_{2y})^2 + (E_{1y} +
E_{2x})^2\right) - \frac{1}{2} K_3 \left(E_{3x}^2 +
E_{3y}^2\right).\label{eq:phas} 
\end{eqnarray}
The last term in (\ref{eq:phas}) is present if the height component in
the third commensurate direction is `rough'. This can be checked by
calculating the finite size corrections to the entropy. For a critical
system these go as 
\begin{equation}
\sigma_{\infty} = \sigma_N - \frac{\pi c v_s}{6 N^2} + {\mathcal O}
\left(\frac{1}{N^2} \right),
\end{equation} 
where $c$ is the central charge and $v_s$ is a geometrical factor. The
central charge measures the number of independent height components in
these kinds of models. Indeed, we find $c=3$. 

Although Eqs. (\ref{eq:BA1})-(\ref{eq:BA4}) in principle can be solved
numerically for very large values of $N$ we found that the structure
of the solution for the groundstate is numerically complicated, which
is why we as yet cannot push the systemsize $N$ larger than 89. This
also has prevented us so far to give estimates for the elastic constants.

\section*{Acknowledgment}
We thank Stichting FOM, which is part of the Dutch foundation for
scientific research NWO.

\section*{References}

\end{document}